\RequirePackage{amsmath}           

\documentclass[
a4paper,                
10pt,                   
conference,             
romanappendices,        
]{IEEEtran} 

\makeatletter
\newcommand{\dontusepackage}[2][]{%
	\@namedef{ver@#2.sty}{9999/12/31}%
	\@namedef{opt@#2.sty}{#1}}
\makeatother

\dontusepackage[tocindentauto]{tocstyle}
\newcommand{\usetocstyle}[1]{}
\newcommand{\settocstylefeature}[2]{}

\usepackage{float}                
\usepackage[backend=biber,
style=numeric-comp,
maxnames=6,
maxcitenames=6,
maxbibnames=6,
doi=true,
isbn=false,
url=false,
sorting=none
]{biblatex}
\IfFileExists{global-do-not-edit.bib}{\addbibresource{global-do-not-edit.bib}}{}

\usepackage{acro}
\robustify\footnote%
\robustify\url%

\makeatletter\newif\ifnewacro%
\@ifpackagelater{acro}{2015/08/15}{
	\setboolean{newacro}{true}
}{
	\setboolean{newacro}{false}
	\typeout{warning: your acro package is too old (<2.0)}
}%
\makeatother
\ifnewacro
\else

\fi

\usepackage{footmisc}

\usepackage{tikz}

\usepackage{hyperxmp}                 

\usepackage{cleveref}                 

\IfFileExists{global-do-not-edit.authors.tex}{\input{global-do-not-edit.authors.tex}}{} 

\author{
	\IEEEauthorblockN{\\
		\parbox{\linewidth}{\centering
			Wenzel Pilar von Pilchau\IEEEauthorrefmark{1},
			Varun Gowtham\IEEEauthorrefmark{6},
			Maximilian Gruber\IEEEauthorrefmark{3},
			Matthias Riedl\IEEEauthorrefmark{2},
			Nikolaos-Stefanos Koutrakis\IEEEauthorrefmark{5},
			Jawad Tayyub\IEEEauthorrefmark{4},
			Jörg Hähner\IEEEauthorrefmark{1},
			Sascha Eichstädt\IEEEauthorrefmark{3},
			Eckart Uhlmann\IEEEauthorrefmark{5},
			Julian Polte\IEEEauthorrefmark{5},
			Volker Frey\IEEEauthorrefmark{4} and
			Alexander Willner\IEEEauthorrefmark{6}
		}
	}
	\and
	\IEEEauthorblockA{\IEEEauthorrefmark{1}University of Augsburg, Augsburg, Germany,\\Organic Computing Group}
	\and
	\IEEEauthorblockA{\IEEEauthorrefmark{3}PTB, Berlin, Germany,\\Presidential Staff}
	\and
	\IEEEauthorblockA{\IEEEauthorrefmark{5}Fraunhofer IPK, Berlin, Germany,\\Production Machines and System Management}
	\and
	\IEEEauthorblockA{\IEEEauthorrefmark{2}ifak e.V., Magdeburg, Germany,\\ICT \& Automation}
	\and
	\IEEEauthorblockA{\IEEEauthorrefmark{4}Endress + Hauser, Maulburg, Germany,\\Technology Development}
	\and
	\IEEEauthorblockA{\IEEEauthorrefmark{6}Fraunhofer FOKUS, Berlin, Germany,\\Open Communication Systems}
	
} 

\newcommand{\metaThanks}{Research for this paper %
	was financed by the Federal Ministry of Education and Research (BMBF) project %
	FAMOUS\footnote{\url{http://famous-project.eu/}}.
	We thank our project partners for their contributions and their collaboration to this research work.}

\DeclareAcronym{AAS}{short={AAS}, long={Asset Administration Shell}}
\DeclareAcronym{Bosch}{short={Bosch}, long={Robert Bosch GmbH}}
\DeclareAcronym{CPS}{short={CPS}, long={Cyber Physical Systems}}
\DeclareAcronym{DT}{short={DT}, long={Digital Twin}}
\DeclareAcronym{E+H}{short={E+H}, long={Endress+Hauser SE \& Co KG}}
\DeclareAcronym{EC}{short={EC}, long={Edge Computing}}
\DeclareAcronym{FAMOUS}{short={FAMOUS}, long={AAS-based modeling for the analysis of mutable CPS}}
\DeclareAcronym{Greedy Computing}{short={Greedy Computing}, long={Greedy Computing}, cite={Foster:2004}}
\DeclareAcronym{I40}{short={I4.0}, long={Industrie 4.0}}
\DeclareAcronym{IIoT}{short={IIoT}, long={Industrial Internet of Things}}
\DeclareAcronym{IPK}{short={IPK}, long={Fraunhofer Institute for Production Systems and Design Technology}}
\DeclareAcronym{IT}{short={IT}, long={Information Technology}}
\DeclareAcronym{ML}{short={ML}, long={Machine Learning}}
\DeclareAcronym{NMI}{short={NMI}, long={National Metrology Institute}}
\DeclareAcronym{OC}{short={OC}, long={Organic Computing}}
\DeclareAcronym{OT}{short={OT}, long={Operation Technology}}
\DeclareAcronym{OPC}{short={OPC UA}, long={Open Platform Communications United Architecture}}
\DeclareAcronym{SI}{short={SI}, long={Système International}}
\DeclareAcronym{API}{short={API}, long={Application Programming Interface}}
\DeclareAcronym{PTP}{short={PTP}, long={Precision Time Protocol}}

\IfFileExists{global-do-not-edit.acro.tex}{\input{global-do-not-edit.acro.tex}}{} 
\AtBeginDocument{}

\newcommand\copyrighttext{%
  \footnotesize \textcopyright 2020 IEEE. Personal use of this material is permitted. Permission from IEEE must be obtained for all other uses, in any current or future media, including reprinting/republishing this material for advertising or promotional purposes,creating new collective works, for resale or redistribution to servers or lists, or reuse of any copyrighted component of this work in other works.
  }
\newcommand\copyrightnotice{%
\begin{tikzpicture}[remember picture,overlay]
\node[anchor=south,yshift=10pt] at (current page.south) {\fbox{\parbox{\dimexpr\textwidth-\fboxsep-\fboxrule\relax}{\copyrighttext}}};
\end{tikzpicture}%
}

\begin{document}

    \title{An Architectural Design for Measurement Uncertainty Evaluation in Cyber-Physical Systems}
    \maketitle
    \copyrightnotice{}

    \begin{abstract}
Several use cases from the areas of manufacturing and process industry, require highly accurate sensor data. As sensors always have some degree of uncertainty, methods are needed to increase their reliability. The common approach is to regularly calibrate the devices to enable traceability according to national standards and \ac{SI} units - which follows costly processes. However, sensor networks can also be represented as \ac{CPS} and a single sensor can have a digital representation (Digital Twin) to use its data further on.
To propagate uncertainty in a reliable way in the network, we present a system architecture to communicate measurement uncertainties in sensor networks utilizing the concept of Asset Administration Shells alongside methods from the domain of Organic Computing.
The presented approach contains methods for uncertainty propagation as well as concepts from the Machine Learning domain that combine the need for an accurate uncertainty estimation.
The mathematical description of the metrological uncertainty of fused or propagated values can be seen as a first step towards the development of a harmonized approach for uncertainty in distributed \ac{CPS} in the context of Industrie 4.0.
In this paper, we present basic use cases, conceptual ideas and an agenda of how to proceed further on.
\acresetall%

    \end{abstract}


    \section{Introduction}\label{sec:introduction}
The goal of \ac{I40} is to drive the formation of an automated factory with cooperation from \ac{CPS}. \ac{I40}, in other words, is aiming to close the gap between the contrasting worlds of \ac{OT} and the \ac{IT}. The field of \ac{OT} predominantly addresses the field of manufacturing and process industry concerning the operation of manufacturing assets such as machines and process knowledge and intends to automate such processes through the means of computing systems keeping safety, reliability and economy. The \ac{IT} intends to make \ac{OT} more efficient and transparent by using and adapting concepts of data processing and enable new business models.

To this end, the concept of a \ac{DT} aims to create a virtual representation/twin of a physical asset, such that a \ac{DT} can be used by professionals on the \ac{IT} side to comprehend field level complexities of the machine. 

In a control loop sequence comprising of "sense-decide-actuate", the sensing stage requires that sensors measuring physical properties are reliable and available. The reliability of a sensor can be established by a calibration traceable to a known reference and based on \ac{SI} units. As a result, a measurement uncertainty can be associated with the measured values from that sensor. This is one of the core principles in the field of metrology - measurement science. Establishing reliability of \ac{CPS} thus requires integrating metrological principles in the data life-cycle from the physical measurement to the Digital Twin and the operational decisions in I4.0. This raises several challenges also for metrology institutes, see \cite{PTB_DigiStudie2018}. 

The intent of this paper is to propose a reference architecture for the purpose of representing physical sensing devices as their digital counterparts. We propose a system which:

\begin{enumerate}
    \item Harvests sensor data from calibrated and reliable sensors through edge devices. 
    \item Enriches the data by adding several other non-reliable sensors in order to extend the dimensions of measurement through sensor fusion.
    \item Pre-processes the sensor data to record characteristics for the purpose of building a mathematical model of the sensor.
    \item Transforms the mathematical model to a digital twin of the sensor.
\end{enumerate}
To quantify the quality of features generated by the application of complex procedures to measurement values, the sensor uncertainty needs to be propagated by specialized approaches. In the field of metrology many methods do exist, but so far aren't available for direct and easy application in \ac{I40} and \ac{CPS}. To bridge this gap, methods and sensors will be encapsulated as agents, providing modular and uncertainty-aware functionality that is abstracted from the user. 
Each sensor agent communicates with other agents - forming a multi-agent-system.
From a higher perspective these agents represent a \ac{DT} of the actual sensor network. A standardized communication along the whole supply chain is organized in an \ac{AAS}. 



In this work-in-progress paper we present our state of development alongside the presentation of our ideas and some concepts how to proceed further on. We present an architectural design for the identification and propagation of measurement uncertainties in \ac{CPS}s, especially in sensor networks. Therefore, we use the concepts of \ac{DT}s and \ac{AAS}s and combine them with methods from the domain of \ac{OC}. 
The remainder of the paper is structured as follows.
We give a brief overview of related work in \Cref{sec:relatedwork}. In the subsequent \Cref{sec:main} we present our use cases, a first architectural design and sketch the planned evaluation phase. The paper is closed with a short summary in \Cref{sec:conclusions}.

    \section{Related Work}\label{sec:relatedwork}

In order to place the contribution of the paper in context
  and identify the gap the work is intended to fill,
  we provide a short literature survey.

In the field of \ac{I40}, academic articles propose directions that can be taken to enable digitization of factories through the use of \ac{DT}s and applying the \ac{EC} paradigm. Modeling concepts were proposed to transform manufacturing towards \ac{DT} concepts \cite{qi18_model_cyber_physic_system_digit}. Impact of \ac{EC} in the form of a roadmap for manufacturing based on key performance indicators was analyzed \cite{georgakopoulos16_inter_thing_edge_cloud_comput_roadm_manuf, fognode2016fmec}. The formalization of \ac{DT}s through \ac{EC} paradigms along with cloud computing brings together centralized and decentralized computing \cite{qi19_smart_manuf_servic_system_based}. The suitability of recent developments in \ac{EC} technologies towards realizing a flexible and distributed open manufacturing ecosystems was studied \cite{li18_towar_open_manuf,lu16_parad_shift_smart_manuf_system_archit}.
A survey on the applicability of \ac{DT}s and their applications opens several research opportunities \cite{lu20_digit_twin_driven_smart_manuf}. The importance of reliable sensor devices and synchronized data exchanges has been stressed \cite{thiyagarajan16_data,vazquez04_multi_sensor_system_using_plast,wan08_anshan,petersen07_requir_oil}. Furthermore, for fields that apply or require sensor fusion, there is an increased requirement of not only the reliability of the sensor but also its temporal and measurement uncertainties \cite{kaempchen2003data,huck11_precis, westenberger11_tempor,hong05_scalab_synch_protoc_large_scale}.

Sensor technology has evolved to support multiple use cases. Particularly in \ac{I40}, the use cases of condition monitoring and machine learning are interesting. Condition monitoring approaches have been proposed to monitor industrial assets leading to anomaly detection \cite{shah18_anomal_iiot} and analytics purposes \cite{uhlmann17_smart_wireless,mahony16_adapt,uhlmann17_decen_data_analy_maint_indus}. 
Machine learning methods have been further proposed to forecast performance of industrial assets based on the collected sensor data \cite{uhlmann18_clust_ident_sensor_data_predic,kaempchen2003data}. The trend in \ac{I40} has led to emergence of \ac{CPS} as a key enabler \cite{vogel-heuser13_anfor_cps_aus_sicht_autom,lee08_cyber_physic_system}.

Our approach shows similarities to all the mentioned research topics, but as all of them set their focus on one specific field, our idea concerns the combination of all these things. Namely the incorporation of measurement uncertainty in \ac{CPS}s, especially sensor networks, and the combination of these with \ac{EC} technology and the concept of \ac{DT}s. 

    \section{Status quo}\label{sec:main}

\subsection{Use Cases}\label{sec:usecases}
We identified several use cases from the area of manufacturing as well as the process industry. These uses cases are presented in this section.

\subsubsection{Manufacturing use cases}
Use cases from the \ac{Bosch} are situated in the ARENA2036\footnote{\url{https://www.bosch.com/research/know-how/success-stories/arena2036/}} and comprise the field of versatile manufacturing. As this concept describes the changing focus from mass production to the manufacturing of few or even single parts and the fast adoption to changing requirements (e.g. yield), machines are getting used in different production chains, depending on the actual demand. To assign resource consumption (energy, $\text{CO}_2$, air pressure, etc.) to single parts, corresponding sensors need to be distributed over the whole production chain(s) to collect appropriate data. The collected data (in combination with the already gathered knowledge) can then be used to get insights of the condition of the machines (condition monitoring). Further on, the time a single part needs for its production can be assigned to it in an accurate way. In the end all of these use cases should be combined with a concluding process result monitoring (e.g. pictures or video stream, audio, vibration, etc.) to achieve the most accurate production cost estimation of a single piece (in different dimensions, e.g. money, power, $\text{CO}_2$, etc.).

As the sensors are distributed along the whole production chain, there arise several challenges that need to be addressed:
\begin{enumerate}
    \item \textbf{Time synchronization}: 
    As parts are identified to be on one machine for a concrete time, the sensor data from exactly this time needs the be related to the corresponding sensors.
    \item \textbf{Uncertainty propagation}: Every sensor has its own uncertainty, but if measurement values are combined and propagated along the production chain, it needs to be assured, that these uncertainties are reliable.
\end{enumerate}

\subsubsection{Process industry use cases}

The Use Cases of \ac{E+H} are situated in two distributed test facilities simulating real process industry environments. 
As digitization and \ac{IIoT} more and more finds its way into the process industry, so far unsolved problems like improvement of capacity without capex, optimization and output of the processes and increase of the productivity through predictive maintenance of sensor devices and assets surrounded by sensor swarms 
are coming into focus. 
In the process industry a certain number of sensors is already installed for the purpose of controlling the process and secure the safety. 
The data of these sensors will be channeled through or bypassed around the control system. 
With the enrichment of data from additional (monitoring) sensors not dedicated to the control of the process we can generate a Digital Twin of the physical sensor network in the digital world. 
By utilization of \ac{ML} and \ac{OC} it is planned to generate higher information out of the network which cannot be derived from the data of the single sensors or out of the control system.
A major question also will be the calculation of the measurement accuracy of a sensor swarm and the measurement information in the \ac{IT} world. 
A possible use case is the generation of a more abstract insight into the process then the bare measurement values given by a single sensor. 
New models for predictive maintenance of sensors based on the data of the swarm and predictive maintenance information for active assets surrounded by a swarm of sensors will be a field of research.

Similar to the use cases from the manufacturing, we face several challenges:
\begin{itemize}
    \item \textbf{Time synchronization}: A swarm of different sensors that combine its data raises the need for an accurate time synchronization.
    \item \textbf{Uncertainty propagation}: As the sensors can be distributed along the whole process, and the data can traverse several fusion operations, it is necessary to have reliable propagation mechanics for the corresponding uncertainty.
\end{itemize}


\subsection{Approach}
The proposed approach to accomplish the defined goals splits into two main blocks: (1) uncertainty and (2) reference model.

\subsubsection{Uncertainty}
It is of interest to enable metrological traceability of measurement values and derived quantities in the emerging field of \ac{IIoT}. 
Although calibrations by an accredited laboratory could provide this kind of information, it is often neglected in the data acquisition and data processing of current \ac{IIoT} systems \cite{eichstadt_2019}.

We propose to enhance measured sensor values with the available (dynamic) calibration information - e.g. from a digital calibration certificate - to provide an uncertainty value for every measured sensor value. 
The combination of an incoming stream of measurement data and the calibration information will take place within the \ac{DT} of the sensor. 
Metrological data processing in successive processing steps is then enabled by requesting the enriched sensor data (value, uncertainty, quantity, unit) from the \ac{DT}. 
Moreover, to ease the correct metrological data processing, methods for common sensor fusion operations need to be developed/provided. 
These methods are required to be in line with the uncertainty propagation according to \cite{jcgm_2008}. 
Examples for sensor fusion operation are: averaging, low-pass-filtering, data-labeling with propagated uncertainty of the label and high quality virtual sensor from multiple lower quality sensors.


Because of the high number of sensors used in \ac{IIoT} systems, information redundancy between sensors is expectable. 
The redundant information can be exploited to overcome another issue tied with high sensor counts -- costly and time-consuming calibration of all installed sensors. 
We therefore want to develop methods to re-calibrate sensors "in the field" in compliance with \ac{NMI}-standards.

\subsubsection{Reference Model}
\begin{figure*}[ht]
	\centering
	\includegraphics[width=0.9\columnwidth]{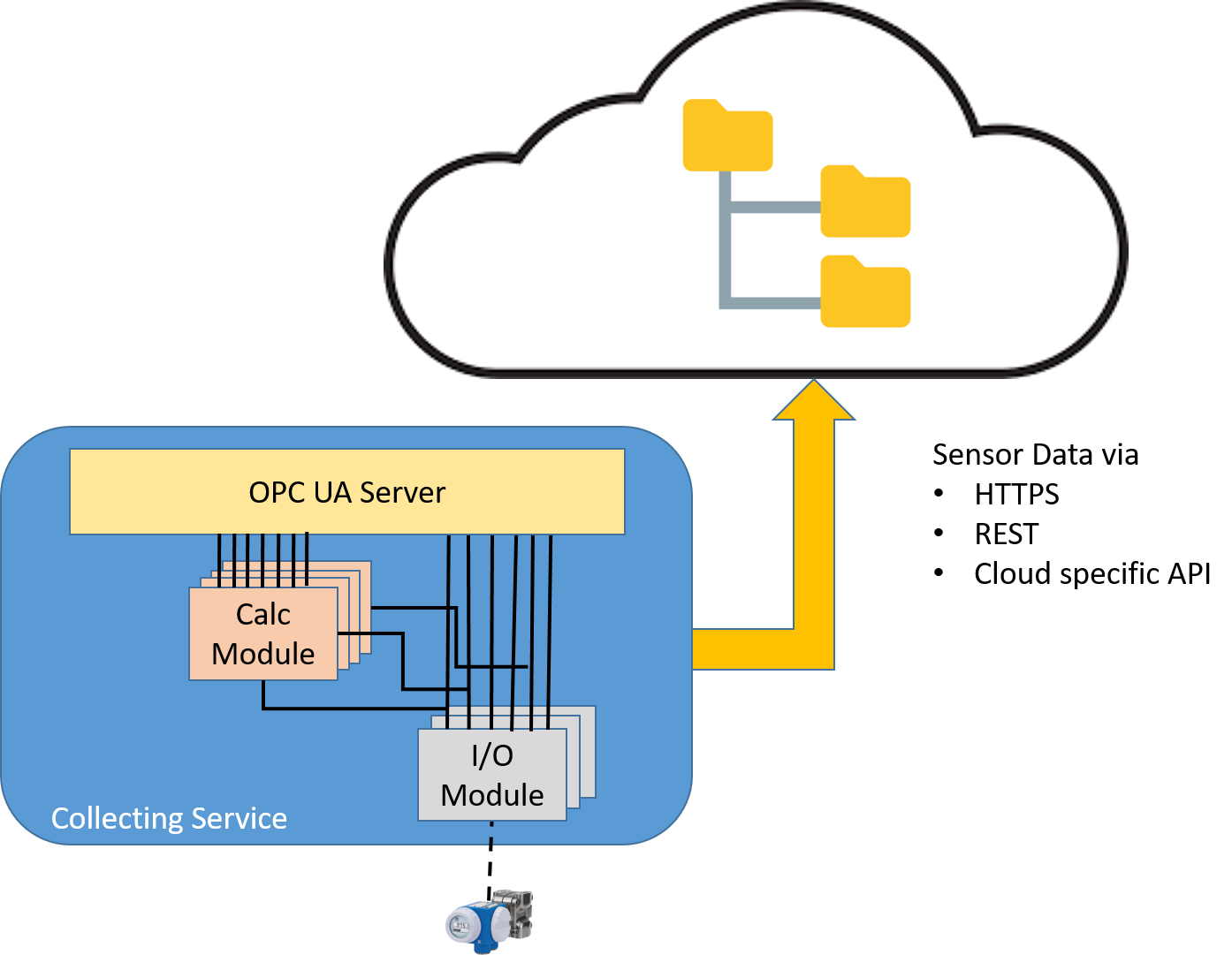}
	\caption{Data collection from field level}
	\label{fig:architecture}
\end{figure*}

The architectural design takes into account the fact that the sensors are installed in field devices, which are connected to the control system via a communication network. 
This can be designed as an industry-specific fieldbus system such as PROFIBUS for Process Automation, HART (wired or wireless) or it can already be based on IT solutions.
It is planned that a collection service (see Fig.~\ref{fig:architecture}) will take up the data from the field level. The implementation can base on a distributed system as described in \cite{8927834} or on a monitoring and supervisory control platform such as ifakFAST\footnote{\url{https://github.com/ifakFAST}\label{fast}}. 
Fieldbus-specific services are triggered in the configured time interval in which the process values are to be provided. 
In addition to the process values, specific attributes such as unit, measuring range adjustment or standardization can be queried.
In the data collector, the recorded values are provided with time stamps. If several data collectors are used on different computer nodes in a system, the local times of the computer nodes are synchronized with each other via the \ac{PTP}, so that further processing takes place on a uniform time basis. 
Virtual sensors can be created in the data collector based on certain rules and provide newly calculated virtual sensor values in the intended cycle.   

Higher level processing is to be based on \ac{AAS}. 
For this purpose, the recorded measured values and their time stamps are mapped into the sub-model of an \ac{AAS} \cite{8868994}, which still has to be defined. 
The implementation is done by mapping the \ac{AAS} model into an \ac{OPC} information model. 
In this way, the evaluation can access online data or evaluate recorded data offline. 
In parallel, the data can also be transferred very specifically to cloud \acp{API}. 

A \ac{DT} can either comprise a single device or a sensor network and to ease its configuration the \textit{Observer-Controller-Architecture} \cite{Prothmann2011} from the \ac{OC} domain is used. 
This highly flexible approach is split into two main parts: the Observer and the Controller. 
As the former can be used for data pre- and/or post-processing, the latter acts as an interface to react onto observed data (e.g. \ac{ML}). 


\subsection{Evaluation}
We plan an evaluation in two phases:

Phase 1 will serve as a first evaluation in a safe environment to identify errors and test the general approach for correctness. 
To ensure the given requirements, the first phase will entirely take place in simulations. Possible simulation frameworks are: DOME\footnote{\url{https://www.ifak-ts.com/pf/ifak-dome/}}, ifakFAST, Assets2036\footnote{\url{https://github.com/boschresearch/assets2036-submodels}}, OpenAAS\footnote{\url{https://acplt.github.io/openAAS/}}, BaSys40\footnote{\url{https://www.basys40.de/}} and Met4FoF\footnote{\url{https://zenodo.org/record/3404800}}. 
One or more of these frameworks will be used depending on the corresponding use cases and an ongoing utility study. 

Phase 2 will take place in the real world. 
Therefore, we build concrete demonstrators at: \ac{Bosch}, \ac{E+H} and \ac{IPK}.
These implementations of the use cases will generate real data for the methods and provide a direct feedback for them. 
Phase 2 is also split in two parts, the first evaluation round comprises the installation of the demonstrators as well as the testing of the developed approaches. 
After the first round, gathered knowledge will be used to improve the architecture and the methods. 
A second round of evaluation will then test the newer versions.

    \section{Conclusion and Future Work}\label{sec:conclusions}




This paper presents our approach to the challenges that arise within sensor networks regarding uncertainty of sensor measurements. 
In collaboration with \ac{Bosch} and \ac{E+H} several use cases from the fields of discrete manufacturing and the process industry were identified.
The use cases cover the monitoring of resource consumption, the generation of high level metrics using raw data, the occurrence of drifts in a sensor network and predictive maintenance.
To achieve the objectives set in the use cases, we propose an approach comprising of a method to manage uncertainty using metrological traceability and a reference model for the architectural design utilizing concepts such as \ac{DT}s and \ac{AAS}'. 
Furthermore, the architecture will be implemented using methods of \ac{OC}. 
To evaluate our concept we intend to initially simulate our approach and ultimately build testbeds at \ac{Bosch}, \ac{E+H} and \ac{IPK}. The objective is to raise the level of metrics generated by sensors and ensure their reliability. 
The presented work will provide potential benefits for industrial users. 



    \section{Acknowledgments}\label{sec:ack}
\metaThanks%

    \printbibliography

\end{document}